\documentclass[twocolumn,aps,prl,groupedaddress,showpacs,slide,nofootinbib]
              {revtex4}%
\usepackage{epsfig}
\begin{document}
\def\bea{\begin{eqnarray}}
\def\eea{\end{eqnarray}}
\def\nn{\nonumber}
\newcommand{\snu}{\tilde \nu}
\newcommand{\sll}{\tilde{l}}
\newcommand{\asnu}{\bar{\tilde \nu}}
\newcommand{\stau}{\tilde \tau}
\newcommand{\dmsnu}{{\mbox{$\Delta m_{\tilde \nu}$}}}
\newcommand{\mt}{{\mbox{$\tilde m$}}}

\renewcommand\epsilon{\varepsilon}
\def\calM{{\cal M}}
\def\be{\begin{eqnarray}}
\def\ee{\end{eqnarray}}
\def\lla{\left\langle}
\def\rra{\right\rangle}
\def\za{\alpha}
\def\zb{\beta}
\def\lsim{\mathrel{\raise.3ex\hbox{$<$\kern-.75em\lower1ex\hbox{$\sim$}}} }
\def\gsim{\mathrel{\raise.3ex\hbox{$>$\kern-.75em\lower1ex\hbox{$\sim$}}} }
\newcommand{\Rbs}{\mbox{${{\scriptstyle \not}{\scriptscriptstyle R}}$}}

\draft


\title{Correlation between lepton flavor violation and $B_{(d,s)} - \overline{B}_{(d,s)}$ mixing\\
in SUSY GUT}

\textwidth 18cm
\textheight 24.5cm

\thispagestyle{empty}
\author{ Kingman Cheung
\thanks{ E-mail : cheung@phys.nthu.edu.tw}}
\affiliation{ Department of Physics and NCTS, National Tsing Hua Univeristy,
     Hsinchu, Taiwan }
\author{ Sin Kyu Kang
\thanks{ E-mail : skkang1@sogang.ac.kr}}
\affiliation{ School of Liberal Arts, Seoul National University of Technology,
       Seoul 139-743, Korea }
\author{ C. S. Kim
\thanks{ E-mail : cskim@yonsei.ac.kr}}
\author{ Jake Lee
\thanks{ E-mail : jilee@cskim.yonsei.ac.kr}}
\affiliation{ Department of Physics, Yonsei University, Seoul
120-749, Korea}

\date{\today}

\begin{abstract}
\noindent
Motivated by the recent measurements of the $B_s-\overline{B}_s$ mass
difference from the D\O\ and CDF  collaborations,
we probe new physics effects in the $B_q-\overline{B}_q$ mixing within the context
of the supersymmetric grand unified model (SUSY GUT). We find that new physics effects
in $B_{s(d)}-\overline{B}_{s(d)}$ mixing lead to the correlated  information in the branching
fractions of the lepton flavor violating decays, which  may serve as a test of
the SUSY GUT. We also discuss the implication of such new physics effects
on the quark-lepton complementarity in the neutrino mixings.

\end{abstract}
\pacs{ ~~~~ Keywords : }
 \maketitle \thispagestyle{empty}
%

\section{Introduction}

The recent measurements for the $B_s-\overline{B}_s$ mass difference from the D\O\ \cite{D0}
and CDF \cite{CDF} collaborations given by
\bea
\label{Bsmix}
 17\,{\rm ps}^{-1} < \Delta M_s^{\rm exp} < 21\,{\rm ps}^{-1}
  ~~~\mbox{(90\%\,CL,~ D\O)}\,, \nn\\
  \Delta M_s^{\rm exp} = (17.77 \pm 0.10 \pm 0.07)\, {\rm ps}^{-1} ~~~\mbox{(CDF)},
\eea
have triggered to probe new physics effects in $b\rightarrow s$ transition.
Although the experimental results are consistent with prediction of the
standard model (SM) dominated by the $t-$quark exchange in the
$B^0_s-\overline{B}^0_s$ box diagram,
they do not fully exclude all the possible new physics effects in $\Delta B=2 $ transitions.
Since flavor changing $b\longrightarrow s$ transitions are
very sensitive to new physics, it is worthwhile to probe them through
the $B_s-\overline{B}_s$ mixing phenomena \cite{bs-works,Ball06,dutta,khalil2}.
Moreover, the combined analysis of the $B_s-\overline{B}_s$ mixing and the $B_d-\overline{B}_d$ mixing
may provide clearer hint on the existence of new physics in flavor changing transitions,
due to a possible cancellation  of hadronic uncertainties.

Probing such a possibility of new physics in flavor changing transitions is
the main purpose of this work.
As a concrete example of physics beyond the SM, we will consider
the supersymmetric grand unified model (SUSY GUT)
with heavy right-handed neutrinos where the imprint of large atmospheric neutrino
mixing may appear in the squark mass matrices.
In this model large Dirac neutrino Yukawa couplings can induce large off-diagonal
mixing in the right-handed
down type squark mass matrix, and there can exist possible correlation between quark
flavor changing processes and
lepton flavor violating (LFV) decays, $\tau\longrightarrow \mu (e) \gamma$.
Here we want to test such new physics in $b\longrightarrow s (d)$ processes in the SUSY GUT,
to find out the correlated new physics effects in LFV decays as a function of
branching fractions of $\tau\longrightarrow \mu (e) \gamma$.
We examine in detail how new physics effects in $B_{s(d)}-\overline{B}_{s(d)}$
mixing phenomena are correlated
with the branching fractions of the LFV decays.
Comparing our analysis with
future measurements of the branching fractions of LFV decays
may serve as a test of the supersymmetric grand unified model.
Recently, the similar idea has been proposed in Ref. \cite{dutta}:
The authors have studied a correlation between $B_s$ mixing and LFV decay
$\tau \longrightarrow \mu \gamma$ by assuming that there is no new physics effects in $B_d$ mixing
and some arbitrary mixing term in the slepton mixing matrix.
However, in our approach,
we consider not only such a correlation but also the correlation
among the ratio of new physics contributions
to $B_d$ and $B_s$ mixings and
the corresponding ratio of the branching fractions of LFV decay $Br(l_i\longrightarrow l_j\gamma)$.
Thus, our approach is less dependent on arbitrary SUSY input parameters
and may also reduce the hadronic uncertainties in those processes.
Moreover, interestingly enough, our analysis can lead to an implication
of new physics effects in the quark-lepton complementarity \cite{qlc} between the solar neutrino
mixing angle and Cabibbo angle, as will be shown later.

In general, the $B^0_q-\overline{B}^0_q$ mass difference is defined as
$\Delta M_{q} = 2 |{\cal M}_{12}(B_q)| \equiv 2 \vert \langle B^0_q \vert
H_{\rm{eff}}^{\Delta B=2} \vert \overline{B}^0_q\rangle \vert$,
where
$H_{\rm{eff}}^{\Delta B=2}$ is the effective Hamiltonian responsible
for the $\Delta B=2$ transition, and the SM prediction \cite{sm-b0b0} is given by
\bea
\calM_{12}^{\rm SM}(B_q)=
\frac{G_F^2m_W^2}{12\pi^2}M_{B_q}\hat{\eta}^B\hat{B}_{B_q}
f^2_{B_q}(V^{\ast}_{tq}V_{tb})^2L_0(x_t),
\eea
where
$G_F$ is the Fermi constant,
$x_t=m_t^2/m_W^2$, $L_0$ is an ``Inami-Lim'' function \cite{ilf},
$\hat{\eta}^B$ is a short-distant QCD correction,
and $f_{B_q}$ and $\hat{B}_{B_q}$ are non-perturbative parameters
from which main theoretical uncertainties arise.

Due to the hadronic uncertainties in the SM prediction for $\calM_{12}^{\rm SM}(B_q)$,
various estimates of the SM values of $\Delta M_{q}^{\rm SM}$ have come out so far consistent.
In order to estimate the SM values of $\Delta M_{q}^{\rm SM}$, in particular,
we adopt the following two results for the input hadronic parameters,
$\hat B_{B_{d,s}} f^2_{B_{d,s}}$.
The first one is from the most recent (unquenched) simulation by  JLQCD
collaboration\,\cite{JLQCD}, with non-relativistic $b$ quark and two
flavors of dynamical light quarks. The second one is from combined
results, denoted by (HP+JL)QCD.
Lacking any direct calculation of $\hat B_{B_q}$ with three dynamical flavors,
it has been suggested to combine the results of $f_{B_q}$ from HPQCD
collaboration\,\cite{HPQCD} with that of  $\hat B_{B_q}$ from JLQCD.
Then, two numerical results for $\Delta M_{q}^{\rm SM}$  are given by \cite{Ball06}
\bea
\label{eq:DMd:SM}
\Delta M_d^{\rm SM}
   &=& \big[ 0.52^{+0.21}_{-0.19} \big]\,{\rm ps}^{-1} \quad {\rm JLQCD} \\ \nn
   &=& \big[ 0.69 \pm 0.14 \big]\,{\rm ps}^{-1}\quad {\rm (HP+JL)QCD}\,,
\eea
\bea
\label{eq:DMs:SM}
\Delta M_s^{\rm SM}
   &=& \big[ 16.1 \pm 2.8 \big]\,{\rm ps}^{-1}\quad {\rm JLQCD} \\ \nn
   &=& \big[ 23.4 \pm 3.8 \big]\,{\rm ps}^{-1}\quad {\rm (HP+JL)QCD} \,.
\eea
The experimental result for $\Delta M_{d}^{\rm exp}$ \cite{bd} is known to be
\bea
\Delta M_{d}^{\rm exp} &=&(0.507\pm 0.004)~{\rm ps}^{-1}.
\eea

The mixing amplitude  including new physics contributions
can be parameterized in a model independent way as
\bea
\label{NPBS}
\calM_{12}(B_q) &=& \calM_{12}^{\rm SM}(B_q)\left[1+R_q\right] \nn \\
        &=&\calM_{12}^{\rm SM}(B_q)\left[1+r_qe^{i\sigma_q}\right]. \label{M12}
\eea
{}From Eq.~(\ref{NPBS}), we obtain
\bea
|R_q|^2 = \left|  \frac{\calM_{12}(B_q)}{\calM_{12}^{\rm SM}(B_q)} \right|^2 + 1
       -2 Re\left(\frac{\calM_{12}(B_q)}{\calM_{12}^{\rm SM}(B_q)}\right).
\eea
Using  $\left|  \frac{\calM_{12}(B_q)}{\calM_{12}^{\rm SM}(B_q)} \right|=
         \frac{\Delta M_{q}^{\rm exp}}{\Delta M_{q}^{\rm SM}}\equiv \Delta_q$,
we get the following relation,
\bea
&&r_q \equiv |R_q|=-\cos\sigma_q\pm\sqrt{\cos^2\sigma_q+\Delta_q^2-1}~,
 \nn \\
\mbox{or}~~~&&\left(\Delta_q -1\right)^2\leq |R_q|^2 \leq \left( \Delta_q +1\right)^2.
\label{relRdelta}
\eea
On the other hand, the complexity of the mixing amplitude $\calM_{12}(B_q)$ leads to the CP violation.
In Eq.~(\ref{M12}),  the
CP phase may be composed of the SM and new physics contributions as
\bea
\phi_q &=& \phi_q^{\rm SM}+\phi_q^{\rm NP}\nn\\
       &=& \phi_q^{\rm SM}+{\rm arg}(1+r_qe^{i\sigma_q}).
\eea
Then, the CP phase arisen from new physics contribution is expressed by
\bea
 \sin \phi^{\rm NP}_{q}=\frac{r_q\sin\sigma_q}
{\sqrt{(1+r_q\cos\sigma_q)^2+(r_q\sin\sigma_q)^2}}~.
\label{newphase}
\eea
{}From the relations Eqs.~(\ref{relRdelta},\ref{newphase}),
we can extract useful information on the new physics
effects in the $B_q-\overline{B}_q$ mixing by using
the experimental results for $\Delta M_{q}^{\rm exp}$ and $\sigma_q$.

\section{SUSY contribution to  $B^0_q-\overline{B}^0_q$ mixing}

As is well known, supersymmetric standard model (SSM) is one of the most motivated candidates
for new physics beyond the standard model.
In supersymmetric theories, the effective Hamiltonian $H^{\Delta B=2}_{\rm eff}$ receives new
contributions through the box diagrams mediated by gluino, chargino, neutralino, and charged
Higgs. It turns out that gluino exchanges give the dominant contributions \cite{susyc}.
It is widely accepted that the mass-insertion approximation in which only terms with off diagonal
elements of squark and slepton mass matrices are considered is profitable to treat the supersymmetric
contributions to flavor changing neutral current processes.
In the mass-insertion approximation, the gluino contribution to the amplitude of $B_s$
oscillation is given in terms of the ratio of the gluino mass $m_{\tilde{g}}$ to the average squark mass
$m_{\tilde{q}}$,
$x \equiv m^2_{\tilde{g}}/m^2_{\tilde{q}}$, and the down squark mass insertions between second and
third generations, $(\delta^d_{AB})_{23}= (m^2_{\tilde d_{AB}})_{23} / m_{\tilde q}^2 $,
where $A$ and $B$ stand for left ($L$) or right ($R$)
handed mixing.

A general expression for $R_s= \calM^{\tilde{g}}_{12}(B_s)/\calM^{\rm{SM}}_{12}(B_s) $
has been given in Ref. \cite{susyc} as follows:
\begin{eqnarray}
R_s&=& a_1(m_{\tilde{q}},x) [(\delta^d_{LL})^2_{23} +
(\delta^d_{RR})^2_{23}] + a_2(m_{\tilde{q}},x)
[(\delta^d_{LR})^2_{23}\nonumber\\
&+& (\delta^d_{RL})^2_{23}] + a_3(m_{\tilde{q}},x)
[(\delta^d_{LR})_{23}(\delta^d_{RL})_{23}] \nonumber\\
&+& a_4(m_{\tilde{q}},x)
[(\delta^d_{LL})_{23}(\delta^d_{RR})_{23}]~, \label{rs}
\end{eqnarray}
where the coefficients $\vert a_1 \vert \simeq {\cal O}(1)$,
$\vert a_2 \vert < \vert a_3 \vert < \vert a_4 \vert \simeq {\cal
O}(100)$.
For example, $\vert a_1 \vert = 7.2$, $\vert a_2 \vert = 129.8$,
$\vert a_3 \vert = 205.7$ and $\vert a_4 \vert = 803.8$
for $m_{\tilde{q}} = 300$ GeV and $x=1$ \cite{khalil2}.
We also obtain similar expression for $R_d= \calM^{\tilde{g}}_{12}(B_d)/\calM^{\rm{SM}}_{12}(B_d) $
by replacing $(\delta^d_{ab})_{23}~(a,b=L,R)$ with $(\delta^d_{ab})_{13}$ in Eq.~(\ref{rs}).
Here, we note that the $LR(RL)$ contributions $(\delta^d_{LR})_{ij}$ are strongly
constrained by the measurement of $b\rightarrow s \gamma$ decay, and thus those
contributions should be very small. {}From now on, we will ignore the $LR(RL)$
contributions to $R_{d,s}$, and thus $R_{s,d}$ is approximately expressed by
\bea
R_s &\simeq & a_1[(\delta^d_{LL})^2_{23}+(\delta^d_{RR})^2_{23}]
 +  a_4[(\delta^d_{LL})_{23}(\delta^d_{RR})_{23}], \nn \\
R_d &\simeq & a_1[(\delta^d_{LL})^2_{13}+(\delta^d_{RR})^2_{13}]
 + a_4[(\delta^d_{LL})_{13}(\delta^d_{RR})_{13}].
 \label{relLLRR}
\eea
We also note that the values of the corresponding
coefficients $a_i$ in $R_d$ are very similar to those in $R_s$.

It was pointed out in \cite{dchang} that large angles in the neutrino sector may
imply large mixing among right-handed down-type quarks if they are grand
unified with lepton doublets, and the imprint of the large atmospheric neutrino
mixing angle may appear in the squark mass matrices as a large
$\tilde{b}_R-\tilde{s}_R$ mixing effect through radiative corrections
due to the large top Yukawa coupling.

In SSM with the right-handed neutrino singlets accounting for the
data on neutrino oscillation, it is known that large neutrino
Dirac Yukawa couplings can induce large off-diagonal mixings in
the right-handed down-type squark mass matrix through
renormalization group evolution. Those mixings can be
parameterized as
$(\delta^d_{RR})_{ij}=(m^2_{\tilde{d}_{RR}})_{ij}/m^2_{\tilde{q}}$,
where $m^2_{\tilde{q}}$ is the average right-handed down-type
squark mass. In particular, the large $\tilde{b}_R-\tilde{s}_R$
mixing generated in turn feeds into new physics effects in $B$
physics, and thus there may be large enhanced $B_s$ mixing. As
shown in SUSY SU(5), the off-diagonal elements in the mass matrix
of the down-type squarks can be generated through RG running, and
is approximately given by \cite{dchang}
\bea
(m^2_{\tilde{d}_{RR}})_{ij}\simeq
-\frac{1}{8\pi^2}(Y^{\dagger}_{\nu}Y_{\nu})_{ij} (3m_0^2+A_0^2)\ln
\frac{M_{\ast}}{M_{\rm GUT}}, \label{lfv}
\eea
where $m_0$ and
$A_0$ stand for the universal scalar mass and the universal
$A$-parameter for the soft SUSY breaking,  $M_{\ast}$ for the
scale where the universality of the scalar mass is imposed
and $M_{\rm GUT}$ denotes the SU(5) breaking scale. Then,
$(m^2_{\tilde{d}_{RR}})_{32}$ and $(m^2_{\tilde{d}_{RR}})_{31}$
contribute to $B_s$ and $B_d$ mixing, respectively.

{}From the experimental results for $\Delta M_{q}$  and their SM predictions,
one can extract the allowed regions of the parameters $|R_q|$ and $\sigma_q$
by using Eq.~(\ref{relRdelta}).
Using these constraints for $|R_q|$, we can extract useful information
on the relevant LFV radiative decays in SUSY GUT context.

\section{Correlation between $B^0_q-\overline{B}^0_q$ mixing and lepton flavor violation}

Let us discuss how the new physics effects extracted from measurements of the $B_q$ mixing can be related
to the lepton flavor violation in the context of SUSY GUT.
In the case that the $RR$ contribution  to $R_q$ dominates over the others in Eq.~(\ref{rs}),
the new physics contributions to $R_q$ are approximately given by
\bea
R_s &\simeq & a_1(\delta^d_{RR})^2_{23}, \nn \\
R_d &\simeq & a_1(\delta^d_{RR})^2_{13}.
\eea
Since the term $(\delta^d_{RR})_{ij}$ is proportional to $(Y^{\dagger}_{\nu}Y_{\nu})_{ij}$ in SUSY GUT,
we can obtain the following simple relation,
\bea
\frac{R_s}{R_d}\simeq \frac{(\delta^d_{RR})^2_{23}}{(\delta^d_{RR})^2_{13}}
\simeq \frac{(Y^{\dagger}_{\nu}Y_{\nu})_{23}^2}{(Y^{\dagger}_{\nu}Y_{\nu})_{13}^2}~.
\label{R-ratio}
\eea
Therefore, the origin of new physics effects in the $B_q$ mixing is from the SUSY seesaw in this case.

On the other hand, the SUSY seesaw model we consider can lead to
sizable effects on the LFV processes such as $l_i\longrightarrow
l_j\gamma $ due to the new source of lepton flavor violation
arisen from the misalignment of lepton and slepton mass matrices,
and the branching ratios of the LFV decays depend on the specific
structure of the neutrino Dirac Yukawa matrix. In the context of
SUSY GUT, this mixing in the charged lepton sector is dictated as
same as that of the down type quark sector in Eq.~(\ref{lfv}). As
discussed in \cite{CKKL}, the LFV processes in SUSY GUT models can
provide a probe of quark-lepton unification. Thus, combining the
idea proposed in \cite{CKKL} with the analysis based on $B_q$
mixing, we can further probe quark-lepton unification. The
contribution to the branching fractions of the LFV decays due to
the slepton mass term is roughly given by \cite{Casas:2001sr} \bea
Br(l_i\rightarrow l_j \gamma ) \simeq
\frac{\alpha^3}{G_F^2}\frac{m^4_{\tilde{q}}}{m_S^8}|(\delta^d_{RR})_{ij}|^2
\tan^2 \beta, \eea where $m_S$ is a supersymmetric leptonic scalar
mass scale and we used a rough GUT relation,
$(\delta^d_{RR})_{ij}\simeq (\delta^l_{LL})_{ij}$. We remark
that such a GUT relationship for the parameter $\delta$'s
must be corrected down to the typical mass scale of the
right-handed Majorana neutrinos $M_R$ because the slepton mass
term gets additional corrections due to RG evolution.
However, in general, such RG effects as well as
corrections from RG evolution down to $M_W$  do not significantly
modify the GUT relations for off diagonal elements
$(\delta^{q,l}_{LL(RR)})_{ij}$ presented at $M_{\rm GUT}$\footnote{
But, there exist highly model dependent
cases with large mixings in neutrino Dirac Yukawa matrices which
may destroy the GUT relation for $\delta$'s due to large change from
RG evolution \cite{ciuchini}. Those cases are not relevant to our
work.}
and, furthermore, the logarithmic scale
dependence is suppressed in the ratio of branching fractions of
LFV processes. 
Relating Eq.~(\ref{R-ratio}) to the expression for
$Br(l_i\rightarrow l_j \gamma ) $, we can derive the following
simple relation \bea \left| \frac{R_s}{R_d} \right| \approx
\frac{Br(\tau\longrightarrow \mu \gamma)}{Br(\tau\longrightarrow e
\gamma)}. \label{relRR1} \eea Therefore, using the experimentally
allowed regions of $|R_s/R_d|$, we can predict the ratio of the
corresponding LFV processes.

However, recent work reported that the SUSY models with the dominant $RR$ mixing
would be disfavored by the $\Delta M_s$ constraints \cite{khalil2}.
In fact, the $LL$ squark mixing receives renormalization group (RG) effects through the CKM matrix.
The evolution from  $M_{\ast}$ to the weak scale $M_W$ leads to the $LL$ mixings such as
\bea
\label{dll23:RG}
(\delta^d_{LL})_{23}\simeq -\frac{1}{8\pi^2}Y^2_tV_{ts}\frac{3m_0^2+A^2}{m_0^2}\ln\frac{M_{\ast}}{M_W}, \\
\label{dll13:RG}
(\delta^d_{LL})_{13}\simeq -\frac{1}{8\pi^2}Y^2_tV_{td}\frac{3m_0^2+A^2}{m_0^2}\ln\frac{M_{\ast}}{M_W},
\eea
where $Y_t$ is the top quark Yukawa coupling, $m_0$ is the typical soft SUSY scale.
It is known that the RG evolution from the GUT scale in supergravity scenario  induces
$(\delta^d_{LL})_{23}\simeq 0.04\sim \lambda^2$ and $(\delta^d_{LL})_{13}\sim \lambda^3$.
Then, the contribution of the double insertion $[(\delta^d_{LL})_{23}(\delta^d_{RR})_{23}]$ in Eq.~(\ref{rs})
should  not be ignored.
In fact, as long as the size of $(\delta^d_{RR})_{23}$ is not greater than $\cal{O}(\lambda)$,
it turns out that the term, $a_4[(\delta^d_{LL})_{23}(\delta^d_{RR})_{23})]$, in Eq.~(\ref{rs})
can dominate over the term, $a_1[(\delta^d_{RR})_{23})]^2$, due to $|a_4|\sim 100 |a_1|$.
Keeping only the leading term, $a_4[(\delta^d_{LL})_{i3}(\delta^d_{RR})_{i3}]$, in $R_{s,d}$,
we obtain the following simple relation between the ratio of $R_s/R_d$ and
the ratio of the branching fractions for LFV decays,
\bea
\left| \frac{R_s}{R_d} \right|^2\approx \left|\frac{V_{ts}}{V_{td}}\right|^2
       \frac{Br(\tau\longrightarrow \mu\gamma)}{Br(\tau\longrightarrow e \gamma)}.
\label{relRR}
\eea
In the following section, we will perform numerical study in detail.
Note that although we keep both contributions proportional to the coefficients $a_1$
and $a_4$ in our numerical analysis, the relation Eq.~(\ref{relRR}) holds reasonably well.

\section{Numerical Estimates and Discussions}

Let us begin by examining how to extract the allowed regions of $r_q$ and $\sigma_q$
based on the SM predictions for $\Delta M_{q}^{\rm SM}$.
As demonstrated in the introduction, we can estimate the SM predictions for $\Delta M_{q}^{\rm SM}$
by using the results of the hadronic parameters from JLQCD collaboration
and the (HP+JL)QCD collaboration.
In this paper, we only take the estimation by the (HP+JL)QCD collaboration as given
by Eqs.~(\ref{eq:DMd:SM}) and (\ref{eq:DMs:SM}), from which we could probe
new physics effects in $B_d$ and $B_s$ mixing more concretely.
As described in Ref. \cite{Ball06}, by using Eqs.~(6-8) and experimental results
for $\Delta M_{q}^{\rm exp}$,
we can obtain new physics contributions
to $\calM_{12}(B_q)$ which is parameterized in terms of $r_q$ and $\sigma_q$.
First, the values of $\Delta_q$ are extracted to be
\bea
\label{Delq:EXP}
\Delta_d &=& 0.75 \pm 0.30 \\
\Delta_s &=& 0.74 \pm 0.18 .
\eea
Using these results for $\Delta_q$ and the relation in Eq.~(\ref{relRdelta}),
one can obtain the constrained regions of the parameters $r_{d,s}$ and $\sigma_{d,s}$, which are
presented in Fig.~1.
In addition,  the parameters $r_d$ and $\sigma_d$ can be further constrained through
the allowed values of the CP phase of new physics in the
$B_d$ mixing $\phi^{\rm NP}_d$ which is obtained by
the experimental value of the $B_d$ mixing phase
$\phi_d (=\phi_d^{\rm SM}(\equiv \beta_{\rm SM})+\phi_d^{\rm NP})$ \cite{Ball06}.
For $B_d$ system, in particular, we have used the known constraint on $\phi_d^{\rm NP}$ given by \cite{Ball06}
\bea
\phi_d^{\rm NP}|_{\rm incl}=-(10.1\pm 4.6)^\circ\,,
\eea
which further constrains $r_{d}$ and $\sigma_{d}$
through the relation of Eq.~(\ref{newphase}).
Note that at present there is no constraint on $\phi_s^{\rm NP}$.

By using the allowed regions of $r_{d,s}$ and $\sigma_{d,s}$ presented in Fig.~1, we can estimate
the value of the ratio $Br(\tau\longrightarrow \mu\gamma)/Br(\tau\longrightarrow e \gamma)$
through the relation Eq.~(\ref{relRR}).
Although it appears that the effects of the phases $\sigma_{d,s}$ vanish away
by taking absolute values in Eq.~(\ref{relRR}),
the phase dependence is still imprinted in the absolute value $|R_q|(=r_q)$,
as shown in Fig.~1.
Please note that the approximate relation Eq.~(\ref{relRR})
holds quite well although it keeps only the contributions proportional to
the coefficient $a_4$ in Eq.~(\ref{relLLRR}).
\begin{figure}[tb]
\vspace*{-5.0cm}
\hspace*{-6cm}
\begin{minipage}[t]{6.0cm}
\epsfig{figure=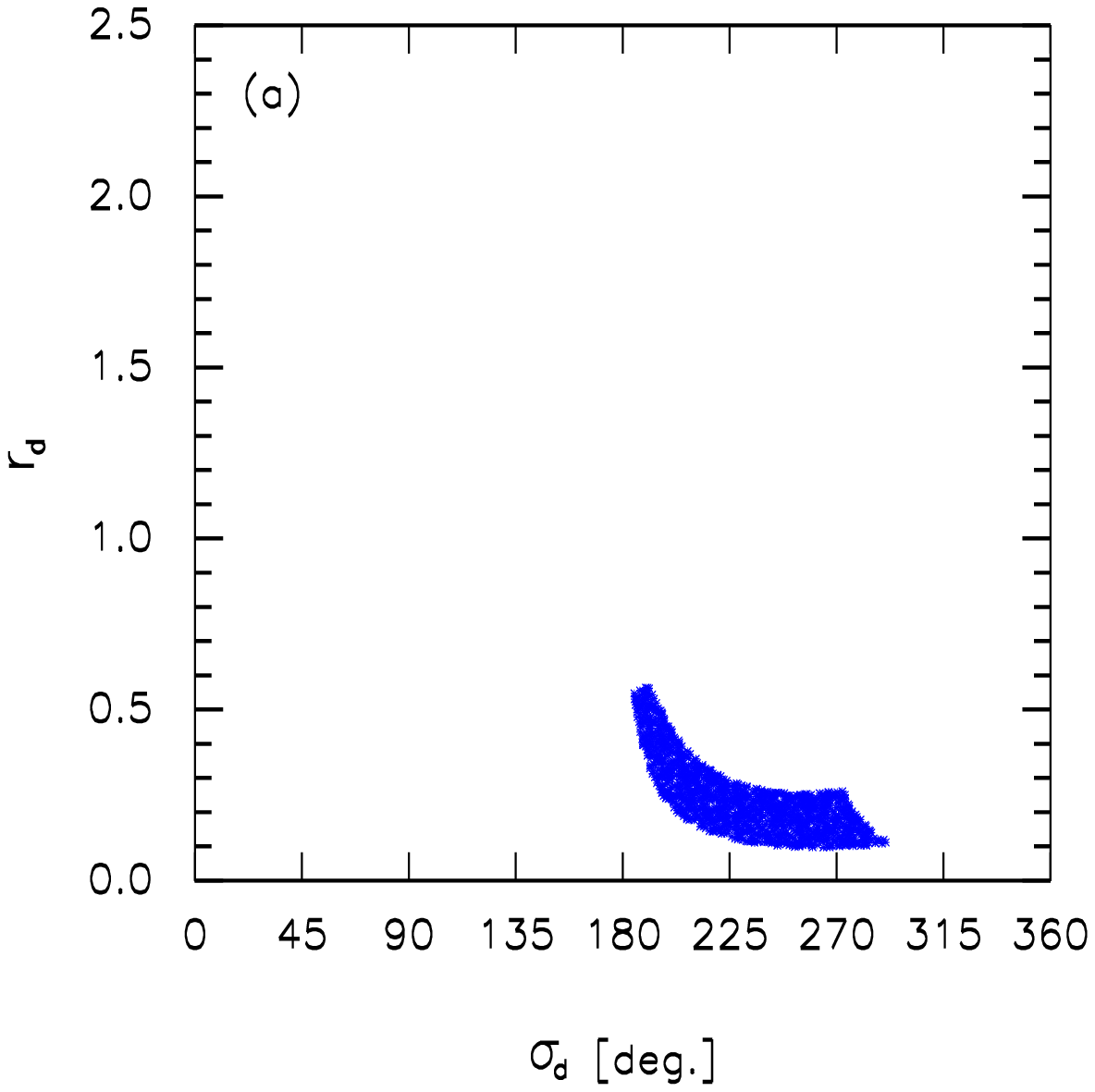,width=13cm,angle=0}
\end{minipage}
\end{figure}
\begin{figure}[tb]
\vspace*{-9.5cm}
\hspace*{-6cm}
\begin{minipage}[t]{6.0cm}
\epsfig{figure=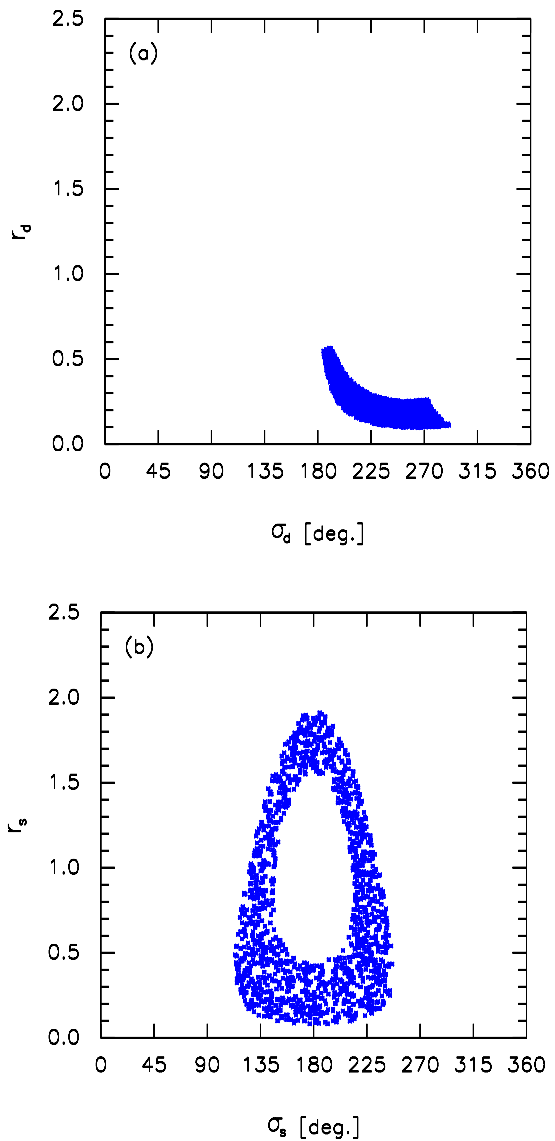,width=13cm,angle=0}
\end{minipage}
\vspace*{-5.3cm}
\caption{\label{Fig1} The constrained region of the parameters $r_{d,s}$ and $\sigma_{d,s}$ obtained
by the method given in Ref. \cite{Ball06} and based on the SM prediction for $\Delta M_{q}^{\rm SM}$
from (HP+JL)QCD collaboration.}
\end{figure}
In Fig.~2, we display the scatter plot of the result for the ratio of branching fractions
$Br(\tau\longrightarrow \mu\gamma)/Br(\tau\longrightarrow e \gamma)$ {\it vs.} the CP phase of new physics
$\phi^{\rm NP}_s$.
Note that the values of the CP phase $\phi^{\rm NP}_s$ are determined in terms of $r_{s}$ and $\sigma_{s}$
through Eq.~(\ref{newphase}).
As shown in Fig.~2, the value of $Br(\tau\longrightarrow \mu\gamma)/Br(\tau\longrightarrow e \gamma)$ decreases
as $\phi^{\rm NP}_s$ approaches zero.
We note that the current limits on LFV radiative decays are $B(\tau\to e\gamma) < 1.1\times 10^{-7}$ and
$B(\tau\to\mu\gamma) < 6.8 \times 10^{-8}$ from BaBar Collaboration \cite{babar},
and $B(\tau\to e\gamma) < 1.2\times 10^{-7}$ and $B(\tau\to\mu\gamma) < 4.5 \times 10^{-8}$
from Belle preliminary report \cite{belle}.
Therefore, if both LFV radiative decays are observed
in near future, we can narrowly constrain the CP phase $\phi^{\rm NP}_s$, and
SUSY GUT, which we now consider, can be ruled out in case that
$Br(\tau\longrightarrow \mu\gamma)/Br(\tau\longrightarrow e \gamma)$
is determined to be below $10^{-3}$.
\begin{figure}[tb]
\vspace*{-5.0cm}
\hspace*{-6cm}
\begin{minipage}[t]{6.0cm}
\epsfig{figure=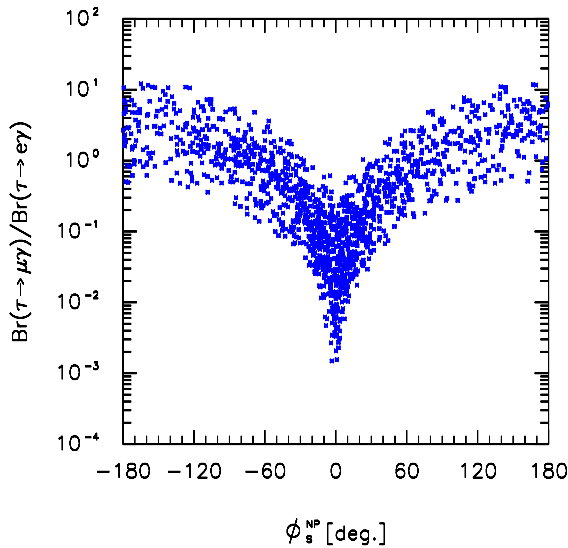,width=13cm,angle=0}
\end{minipage}
\vspace*{-5.3cm} \caption{\label{Fig2} The ratio
$Br(\tau\longrightarrow \mu\gamma)/Br(\tau\longrightarrow e
\gamma)$ {\it vs.} the CP phase of new physics $\phi^{\rm NP}_s$.
}
\end{figure}
Inversely, if we determine the size of $\phi^{\rm NP}_s$
as well as that of $r_{s,d}$ narrowly enough, we can predict the value of
$Br(\tau\longrightarrow \mu\gamma)/Br(\tau\longrightarrow e \gamma)$, and thus
our approach can serve as a test of the SUSY GUT.

We also calculate the branching fraction for the LFV decay, $\tau \rightarrow \mu \gamma$.
In Fig.~3, we present the prediction of $ Br(\tau\longrightarrow \mu\gamma)$ {\it vs.} the CP phase of new physics
$\phi^{\rm NP}_s$.
Here we adopt a universal scalar mass, $m_0=m_S=m_{\tilde q}=300$ GeV, $A_0=0$ and $\tan\beta=10$.
For the scale $M_\ast$ in Eqs.~(\ref{dll23:RG}) and (\ref{dll13:RG}), we take $M_\ast=10^{17}$ GeV.
{}From Fig.~3 with the upper bound of Belle, one can constrain
$-120^\circ < \phi_s^{\rm NP} < 120^\circ$.
However, we note that the prediction for $ Br(\tau\longrightarrow \mu\gamma)$ strongly depends
on the SUSY parameters, whereas the ratio of the branching fractions
is almost independent of the SUSY input parameters.


\begin{figure}[tb]
\vspace*{-5.0cm}
\hspace*{-6cm}
\begin{minipage}[t]{6.0cm}
\epsfig{figure=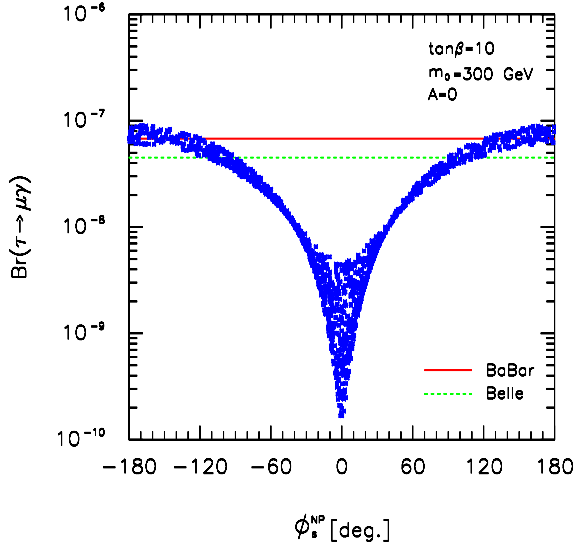,width=13cm,angle=0}
\end{minipage}
\vspace*{-5.3cm}
\caption{\label{Fig3} $ Br(\tau\longrightarrow \mu\gamma)$ {\it vs.} $\phi^{\rm NP}_s$.
Here we adopt a universal scalar mass $m_0=m_S=m_{\tilde q}=300$ GeV, $A_0=0$ and $\tan\beta=10$.
The horizontal lines are
the current upper bounds of BaBar \cite{babar} and Belle \cite{belle} collaborations, respectively.}
\end{figure}

It turns out that our numerical analysis based on Eq.~(\ref{relLLRR}) leads to
\bea
\frac{Br(\tau\longrightarrow \mu \gamma)}{Br(\tau\longrightarrow e \gamma)}<\frac{1}{\lambda^2}~,
\label{resultRR}
\eea
where $\lambda(\simeq 0.22)$ is a parameter of Wolfenstein parametrization of the CKM matrix.
We note that  this upper bound is
almost the same as that obtained from the case with only $RR$ contribution.
The existence of the upper bound on
$\frac{Br(\tau\longrightarrow \mu \gamma)}{Br(\tau\longrightarrow e \gamma)}$
may lead to an implication on the parametrization of neutrino mixing matrix with regard to quark-lepton
unification \cite{CKKL}.
As shown in Ref. \cite{CKKL}, the so-called quark-lepton complementarity relation,
$\theta_{sol}+\theta_C=\pi/4$
between the solar mixing angle $\theta_{sol}$ and Cabibbo angle $\theta_C$,
can be interpreted as a support for
the quark-lepton unification, which in turn makes it possible
to decompose the neutrino mixing matrix into
a CKM-like matrix and a bi-maximal mixing matrix.
It has also been shown that a particular
parametrization of $U_{\rm PMNS}$ with regard to
quark-lepton unification could be singled out by examining the ratios of the branching fractions
$Br(l_i \longrightarrow l_j \gamma)$.
It turns out that  the result given by Eq.~(\ref{resultRR}) may indicate that
the neutrino mixing matrix $U_{\rm PMNS}$ parameterized by
$U^\dag (\lambda) U_{\rm bimax}$ with CKM-like matrix $U(\lambda)$
and the bi-maximal mixing matrix $U_{\rm bimax}$ is preferred.

In summary, motivated by the recent measurements for the $B_s-\overline{B}_s$ mass difference from the D\O\ \cite{D0}
and CDF \cite {CDF} collaborations,
we have probed new physics effects in the $B_q-\overline{B}_q$ mixing in the context of the
supersymmetric grand unified model. We have found that new physics effects
in $B_{s(d)}-\overline{B}_{s(d)}$ mixing lead to the correlated  information on the branching
fractions of the lepton flavor violating decays, which  may serve as a test of
the SUSY GUT.
We have also discussed the implication of such new physics effects
on the quark-lepton complementarity in the neutrino mixings.

\section{acknowledgement}

\noindent KC is supported in part by NSC Taiwan under grant no. 95-2112-M-007-001
and by the NCTS. SKK is supported in part by the KOSEF SRC program through CQUeST with Grant No.
R11-2005-021 and in part by the KRF Grant funded by the
Korean Government(MOEHRD) (KRF-2006-003-C00069).
CSK is supported in part by CHEP-SRC and in part
by the KRF Grant funded by the Korean Government (MOEHRD)
No. KRF-2005-070-C00030.
JL is supported by BK21 program of the Ministry of Education in Korea.

\end{document}